\documentclass[10pt]{article}
\usepackage{amsmath}
\usepackage{cite}
\def\be#1{\begin{equation}\label{#1}}
          \def\ee{\end{equation}}

\usepackage{pdfpages}
\usepackage[normalem]{ulem}

\title{Covariant Mass and Geometrical setup in Euclidean gauge theories}

\author{A.R. de S\'{a}$^{a}$\footnote{email:alineribeirodesa@gmail.com.br}, M.A.L. Capri$^{a}$\footnote{email: marcio@dft.if.uerj.br} ,
V. E. R. Lemes$^{a}$\footnote{email: verlemes@gmail.com.br}, \\
\small \em $^a$Instituto de F\'\i sica, Universidade do Estado do Rio de
Janeiro,\\
\small \em Rua S\~{a}o Francisco Xavier 524, Maracan\~{a}, Rio de Janeiro - RJ,
20550-013, Brazil}

\begin{document}
\maketitle
\begin{abstract}
\noindent A nonlocal mass operator is
consistently defined in the local form through the introduction of a set of additional 
fields with geometrical appropriated properties. A local and polynomial gauge-invariant action is thus established. Equations compatible with the study of renormalization, from the algebraic point of view, are presented in the Landau Gauge.
		\end{abstract}
%%%%%%%%%%%%%%%%%%%%%%%%%%%%%%%%%%%%%%%%%%%%%%%%%%%%%%%%%%%%%%%%%%%%%%%%%%%%%%%%%%%%%%%%%%%%
	\section{Introduction}

Non-maximal dimension condensates are of fundamental interest in order to study the infrared into euclidean Yang-Mills theories as
we can see by the amount of results obtained through
theoretical, phenomenological  and lattice
simulations
\cite
{Cornwall:1981zr,Greensite:1985vq,Stingl:1985hx,Lavelle:1988eg,Gubarev:2000nz,Gubarev:2000eu,
Verschelde:2001ia,Kondo:2001nq, Kondo:2001tm,Dudal:2003vv,
Browne:2003uv,Dudal:2003gu, Dudal:2003by,
Dudal:2004rx,Browne:2004mk, Gracey:2004bk,
Li:2004te,Boucaud:2001st,Boucaud:2002nc,
Boucaud:2005rm, RuizArriola:2004en,Suzuki:2004dw,Gubarev:2005it,Furui:2005bu,Boucaud:2005xn,Chernodub:2005gz}.
One particularly important case is the dimension two operator $A_{\mu }^{a}A_{\mu }^{a}$ which is not gauge invariant but is multiplicatively renormalizable to all orders in the Landau gauge. Other dimension two operators can be renormalizable at a large number of other gauges like linear covariant gauges \cite{Dudal:2003np}, Curci-Ferrari
and the maximal Abelian gauge \cite{Dudal:2003pe, Dudal:2003gu}.

 The fundamental problem of these operators like $A_{\mu }^{a}A_{\mu }^{a}$ and the condensate $\left\langle A_{\mu }^{a}A_{\mu}^{a}\right\rangle $is that they are gauge dependent and all of their quantum properties must be defined in each gauge. Many attempts in order to present a gauge invariant mechanism for these operators can be done by \cite{Delbourgo:1986wz, Delbourgo:1987np, Ruegg:2003ps} but the result is always non unitary, nonrenormalizable or both.
In this way it is a quite natural objective to discuss and present a suitable colorless dimension two operator $%
\mathcal{O}(A)$ which preserves gauge invariance
\begin{eqnarray}
\delta \mathcal{O}(A) &=&0\;,  \nonumber \\
\delta A_{\mu }^{a} &=&-D_{\mu }^{ab}\omega ^{b}\;,  \label{gauge0}
\end{eqnarray}
where $D_{\mu }^{ab}$ is the covariant derivative, given by
\begin{equation}
D_{\mu }^{ab}=\delta ^{ab}\partial _{\mu }-gf^{abc}A_{\mu }^{c}\;.
\label{cov0}
\end{equation}
The natural candidate for this is the nonlocal operator \begin{equation}
\mathcal{O}(A)=-\frac{1}{2}\int d^{4}xF_{\mu \nu }^{a}\left[ \left(
D^{2}\right) ^{-1}\right] ^{ab}F_{\mu \nu }^{b}\;.  \label{gm}
\end{equation} these operator is studied in \cite{Capri:2005dy}.The method developed consists of writing the nonlocal operator into a local form as:
\begin{equation}
\frac{m^{2}}{4}\int d^{4}xF_{\mu \nu }^{a}(\frac{1}{D^{2}})^{ab}F_{\mu \nu }^{b} \Rightarrow \int d^{4}x\left( \frac{1}{4}\bar{B}_{\mu \nu }^{a}D_{\sigma
}^{ab}D_{\sigma }^{bc}B_{\mu \nu }^{c}+\frac{im}{4}\left( B-\bar{B%
}\right) _{\mu \nu }^{a}F_{\mu \nu }^{a}\right). 
\end{equation}
Unfortunately by means of algebraic renormalization methods is possible to observe that a mass counterterm in the tensorial fields is obtained and these counterterm turns localization of the gauge invariant nonlocal mass into an impossibility only with usual antissimetric tensor fields\cite{Capri:2005dy}. In fact the obtained result implies that is necessary to include to the quantum action, terms like
\begin{equation}
-\frac{3}{8}%
m^{2}\lambda _{1}\left( \bar{B}_{\mu \nu }^{a}B_{\mu \nu
}^{a}\right) +
m^{2}\frac{\lambda _{3}}{32}\left( \bar{B}_{\mu \nu }^{a}-B_{\mu
\nu }^{a}\right) ^{2} + \frac{\lambda^{abcd}}{16}\left( \bar{B}_{\mu\nu}^{a}B_{\mu\nu}^{b}%
\right)\left( \bar{B}_{\rho\sigma}^{c}B_{\rho\sigma}^{d}%
\right).
\end{equation}
These terms turn impossible the localization of the gauge invariant nonlocal mass operator with these simple mechanism.It is important to emphasize here that this analysis in no way prohibits a more elaborate mechanism from achieving these goal.
Now we will present a mechanism that could turn possible to localize the nonlocal operator (\ref{gm}) without the possibility of mass terms in the localizing action that can destroys the process.

%%%%%%%%%%%%%%%%%%%%%%%%%%%%%%%%%%%%%%%%%%%%%%%%%%%%%%%%%%%%%%%%%%%%%%%%%%%%%%%%%%%%%%%%%%%%
 \section{Localization of the operator 
 $\int d^{4}xF_{\mu \nu}\frac{1}{D^{2}}F_{\mu \nu }$ with self-dual and anti-self-dual tensor fields.}

First of all it is necessary to provide a geometrical mechanism that could avoid mass terms in the localizing tensorial fields. This geometrical mechanism is recognizable in the self-dual and anti-self-dual property. In simple terms 
two antissimetric tensor fields witch obey the following relation $\overline{\varphi}_{\mu \nu } \varphi^{\mu \nu }=0 $ are necessary in order to avoid mass term in the localizing fields.
The most simple way in order to achieve this goal is to introduce these fields in the localizing action with projectors that has the desired propertie \textit{i.e}.
\begin{eqnarray}
  \overline{\theta }_{\mu \nu \alpha \beta } &=& \frac{1}{4}(\delta_{\mu \alpha }\delta_{\nu \beta  }-\delta_{\mu \beta  }\delta_{\nu \alpha }+\epsilon_{\mu \nu \alpha \beta })\nonumber \\
  \theta_{\mu \nu \alpha \beta }&=& \frac{1}{4}(\delta_{\mu \alpha }\delta_{\nu \beta  }-\delta_{\mu \beta  }\delta_{\nu \alpha }-\epsilon_{\mu \nu \alpha \beta })\nonumber \\
  \overline{\theta }_{\mu \nu \alpha \beta }\overline{\theta }^{\alpha \beta \sigma \lambda }&=& \overline{\theta }_{\mu \nu }^{\,\,\,\,\,\,\sigma \lambda }\nonumber \\
 \theta_{\mu \nu \alpha \beta } \theta^{\alpha \beta \sigma \lambda }&=&\theta_{\mu \nu }^{\,\,\,\,\,\,\sigma \lambda } \nonumber \\
 \overline{\theta }_{\mu \nu \alpha \beta }\theta^{\alpha \beta \sigma \lambda }&=& 0 \nonumber \\
 \epsilon_{\alpha \beta \mu \nu} \epsilon^{\alpha \beta \sigma \lambda } &=&  2\delta_{[\mu \nu ]}^{\sigma \lambda } \nonumber \\
 \delta_{[\mu \nu ]}^{\sigma \lambda } &=& \delta_{\mu }^{\sigma }\delta_{\nu }^{\lambda }-\delta_{\nu }^{\sigma }\delta_{\mu }^{\lambda } \nonumber \\
 \epsilon_{\alpha \beta \mu \nu } \epsilon^{\alpha \epsilon  \sigma \lambda }&=& \delta_{\beta }^{\epsilon }\delta_{[\mu \nu ]}^{\sigma \lambda } + \delta_{\beta }^{\lambda }\delta_{[\mu \nu ]}^{\epsilon \sigma  } +\delta_{\beta }^{\sigma }\delta_{[\mu \nu ]}^{\lambda \epsilon } \label{projectors}
\end{eqnarray}
and 
\begin{eqnarray}
  \overline{\varphi}_{\mu \nu } &=& \overline{\theta }_{\mu \nu \alpha \beta }\overline{\varphi}^{\alpha \beta }\nonumber \\
  \varphi_{\mu \nu } &=& \theta _{\mu \nu \alpha \beta } \varphi ^{\alpha \beta }.
\label{dualfields}
\end{eqnarray}
which leads to the following solution:
\begin{eqnarray}
\overline{\varphi}_{\mu \nu } &=& \overline{T}_{\mu\nu} + \widetilde{\overline{T}}_{\mu\nu}, \;\;\;\;
\widetilde{\overline{T}}_{\mu\nu}= \frac{1}{2}\epsilon_{\alpha \beta \mu \nu }\overline{T}^{\alpha\beta}\nonumber \\
\varphi_{\mu \nu } &=& T_{\mu\nu}-\widetilde{T}_{\mu\nu},\;\;\;\;
\widetilde{T}_{\mu\nu}= \frac{1}{2}\epsilon_{\alpha \beta \mu \nu }T^{\alpha\beta}.
\end{eqnarray}
This solution is spite of being interesting, is not the best form to write the action
in order to explicitly obtain the set of equations compatible with the quantum action
principle. Due to this we will continue with the fields $\overline{\varphi}_{\mu \nu }$ and $\varphi_{\mu \nu }$.
It is important to stress the following property 
\begin{equation}
\overline{\theta }_{\mu \nu \alpha \beta } \theta^{\mu \gamma \lambda \rho  }\neq 0 . 
\end{equation}
This property is fundamental for the construction of a dynamical kynetic term.
It is also is important to note that these two properties are also relevant in the study of infrared properties. The generation of mass for all  components of the gauge field is different from the usual symmetry breaking. This apparent paradox is solved in this model in a different way of the one presented in \cite{Capri:2005dy}.
To understand the mechanism that we will present it is instructive to remember the original mechanism presented in \cite{Capri:2005dy}.
It starts adding the nonlocal mass operator to the Yang-Mills action , \textit{i.e. } considering:
\begin{equation}
S_{YM}+S_{\mathcal{O}}\;,  \label{ymop}
\end{equation}
where
\begin{equation}
S_{YM}=\frac{1}{4}\int d^{4}xF_{\mu \nu }^{a}F_{\mu \nu }^{a}\;,  \label{ym}
\end{equation}
and
\begin{equation}
S_{\mathcal{O}}=-\frac{m^{2}}{4}\int d^{4}xF_{\mu \nu }^{a}\left[ \left(
D^{2}\right) ^{-1}\right] ^{ab}F_{\mu \nu }^{b}\;.  \label{massop}
\end{equation}
The term (\ref{massop}) is localized by means of the introduction of a
pair of bosonic antisymmetric tensor fields in the adjoint
representation, $\left( B_{\mu \nu }^{a},\bar{B}_{\mu \nu }^{a}\right) $,
according to
\begin{eqnarray}
e^{-S_{\mathcal{O}}} &=& \int D\bar{B}DB (\det D^{2})^{6} [-S_{\mathcal{O}L}]  \nonumber \\
S_{\mathcal{O}L}  &=& (\frac{1}{4}\int d^{4}x\bar{B}_{\mu \nu }^{a}
D_{\sigma }^{ab}D_{\sigma }^{bc}B_{\mu \nu }^{c} 
+  \frac{im}{4}\int d^{4}x(B-\bar{B})_{\mu \nu }^{a}F_{\mu \nu }^{a}) ),
\end{eqnarray}
where the determinant, $\left( \det D^{2}\right) ^{6}$, takes into account
the Jacobian arising from the integration over the bosonic fields $%
\left(\bar{B}_{\mu \nu }^{a},B_{\mu \nu }^{a}\right) $. This term can also
be localized by means of suitable anticommuting antisymmetric tensor fields $%
\left( \bar{G}_{\mu \nu }^{a},G_{\mu \nu }^{a}\right) $, namely
\begin{equation}
\left( \det D^{2}\right) ^{6}=\int D\bar{G}DG\exp \left( \frac{1}{4}\int {%
d^{4}x}\bar{G}_{\mu \nu }^{a}D_{\sigma }^{ab}D_{\sigma }^{bc}G_{\mu \nu
}^{c}\right) \;.  \label{loc3}
\end{equation}
The bosonic fields $\left( \bar{B}_{\mu \nu }^{a},B_{\mu \nu }^{a}\right) $  and the anticommuting fields $%
\left( \bar{G}_{\mu \nu }^{a},G_{\mu \nu }^{a}\right) $ form a quartet \cite{Capri:2005dy}. Taking into account that in a certain moment the gauge parameter $\omega$ will be promoted to a ghost $c$ this quartet became a BRST quartet.The complete set of BRST equations for the Localizing fields is given by:
\begin{eqnarray}
\delta A_{\mu }^{a} &=&-D_{\mu }^{ab}\omega ^{b}\;,  \nonumber \\
\delta B_{\mu \nu }^{a} &=&gf^{abc}\omega ^{b}B_{\mu \nu }^{c} + G_{\mu \nu }^{a}\;,  \nonumber
\\
\delta \bar{B}_{\mu \nu }^{a} &=&gf^{abc}\omega ^{b}\bar{B}_{\mu \nu }^{c}\;,
\nonumber \\
\delta G_{\mu \nu }^{a} &=&gf^{abc}\omega ^{b}G_{\mu \nu }^{c}\;,  \nonumber
\\
\delta \bar{G}_{\mu \nu }^{a} &=&gf^{abc}\omega ^{b}\bar{G}_{\mu \nu }^{c} + \bar{B}_{\mu \nu }^{a}\;,
\label{gtm}
\end{eqnarray}
so the Kinetic part of the localizing action \begin{eqnarray}
S_{BG} &=&\frac{1}{4}\int d^{4}x\left( \bar{B}_{\mu \nu }^{a}D_{\sigma
}^{ab}D_{\sigma }^{bc}B_{\mu \nu }^{c}-\bar{G}_{\mu \nu }^{a}D_{\sigma
}^{ab}D_{\sigma }^{bc}G_{\mu \nu }^{c}\right) \;  \label{actionsBG}
\end{eqnarray}
is left invariant but the massive part of the localizing action 
\begin{eqnarray}
S_{m} &=&\frac{im}{4}\int d^{4}x\left( B-\bar{B}\right) _{\mu \nu
}^{a}F_{\mu \nu }^{a}\;.  \label{actions-M}
\end{eqnarray} 
is not invariant.
In order to avoid that problem the mass action $S_{m}$ is replaced by \begin{equation}
\frac{1}{4}\int d^{4}x\left( V_{\sigma \rho \mu \nu }\bar{B}_{\sigma \rho
}^{a}F_{\mu \nu }^{a}-\bar{V}_{\sigma \rho \mu \nu }B_{\sigma \rho
}^{a}F_{\mu \nu }^{a}\right) \;.  \label{rs}
\end{equation}
At the end, the sources $V_{\sigma \rho \mu \nu }(x)$, $\bar{V}_{\sigma \rho
\mu \nu }(x)$ are required to attain their physical value, namely
\begin{equation}
\bar{V}_{\sigma \rho \mu \nu }\Big|_{\mathrm{phys}}=V_{\sigma \rho \mu \nu }%
\Big|_{\mathrm{phys}}\;=\;\frac{-im}{2}\left( \delta _{\sigma \mu }\delta
_{\rho \nu }-\delta _{\sigma \nu }\delta _{\rho \mu }\right) \;,  \label{ps}
\end{equation}
so that expression (\ref{rs}) gives back the term $S_{m}$.
This is essentially a trick in order to treat a non invariant action term in an algebraic renormalization scheme. Here it is necessary to enphasize that these mechanism has many problems from the renormalization point of view. In the algebraic renormalization scheme classical sources can be set to any classical value at any point in the renormalization calculus. We will abandon these mechanism in order to remain strictly into the algebraic renormalization scheme observing that all equations and symmetry identityes are compatible with the quantum action principle \textit{i.e} Ward identities with and without linear breaking and Slavnov equation without breaking. There are three points that still need some attention into these procedure. Firstly we need a geometrical mechanism that can block mass terms like 
\begin{equation}
 \tilde{S}_{m} = \int d^{4}x\left[  m^{2}\left(\bar{B}_{\mu \nu }^{a}B_{\mu \nu }^{a}-\bar{G}_{\mu \nu }^{a}G_{\mu \nu }^{a}\right)\right] . 
\end{equation}
This can be done by introducing tensor fields coupled to the projectors or obeying the equations as presented in (\ref{dualfields}). Again it is important to remember 
that due to (\ref{projectors}) a mass term like the above one is forbidden if the fields  $\left( \bar{B}_{\mu \nu }^{a},B_{\mu \nu }^{a}\right) $ are replaced by $\left( \overline{\theta} _{\mu \nu \alpha \beta } \overline{\varphi }^{a\,  \alpha  \beta  },\theta _{\mu \nu \alpha \beta }\varphi ^{a \,\alpha \beta }\right) $. It is also clear that the anticommunting fields $\left( \bar{G}_{\mu \nu }^{a},G_{\mu \nu }^{a}\right) $ are also substituted by $\left( \overline{\theta} _{\mu \nu \alpha \beta } \overline{\omega  }^{a \,\alpha  \beta  },\theta _{\mu \nu \alpha \beta }\omega  ^{a \,\alpha \beta }\right) $.
Secondly a quantum mechanism that garantee that a source be a well defined classical term and not a specific fixed  mass value\footnote{In order to use algebraic renormalization scheme to ensure the renormalizability of the quantum action it is necessary that all the transformations for a classical source give rise to another classical source\cite{Piguet:1995er}. This only opens the possibility for a symetry breaking mechanism or a linear soft breaking symmetry. } and the third and final point is a way to define two different phases, a massive and a non massive one. In order to solve that problem we will introduce another quartet of scalar fields 
\begin{eqnarray}
\delta \overline{\psi } &=& \overline{\phi } \nonumber \\
\delta \overline{\phi } &=& 0 \nonumber \\
\delta \phi &=& \psi \nonumber \\
\delta \psi &=& 0
\end{eqnarray}
and do a symmetry breaking into these scalar fields $(\overline{\phi },\phi )$.
This mechanism is clearly not exactly a standard one. The fundamental difference is that the scalar fields are not linked to a non-Abelian group and due to this, the spontaneous symmetry breaking mechanism give mass to all components of the non-Abelian gauge field $A_{_{\mu }}^{a}$. Also the two phases are defined in the symmetry breaking mechanism.

The proper action in order to do that is:
\begin{eqnarray}\label{lcma}
S_{YM+OP} &=&\int d^{4}x \,\, \{  \frac{1}{4}F_{\mu \nu }^{a}F_{\mu \nu }^{a} + \overline{\theta }_{\mu \nu \alpha \beta }(D^{\nu }\overline{\varphi}^{\alpha \beta  })^{a}\theta^{\mu \sigma \lambda \rho }(D_{\sigma }\varphi_{\lambda \rho })^{a} \nonumber \\
&-& \overline{\theta }_{\mu \nu \alpha \beta }(D^{\nu }\overline{\omega }^{\alpha \beta  })^{a}\theta^{\mu \sigma \lambda \rho }(D_{\sigma }\omega _{\lambda \rho })^{a} - i\overline{\phi}\varphi_{\mu \nu }^{a}\theta^{\mu \nu \alpha \beta}F_{\alpha \beta }^{a} \nonumber \\
&+& i\phi \overline{\varphi }_{\mu \nu }^{a}\overline{\theta }^{\mu \nu \alpha \beta }F_{\alpha \beta }^{a} 
+ i\overline{\psi }\omega _{\mu \nu }^{a}\theta^{\mu \nu \alpha \beta}F_{\alpha \beta }^{a} 
+ i\psi  \overline{\omega }_{\mu \nu }^{a}\overline{\theta }^{\mu \nu \alpha \beta }F_{\alpha \beta }^{a} \nonumber \\
&+& \partial_{\mu }\overline{\phi }\partial^{\mu } \phi - \partial_{\mu }\overline{\psi }\partial^{\mu } \psi 
- m^{2}(\overline{\phi }\phi  -\overline{\psi }\psi ) +\frac{\lambda }{2}(\overline{\phi }\phi  -\overline{\psi }\psi )^{2} \},
\end{eqnarray}
to which adding the Landau gauge fixing action
\begin{equation}
 S_{gf} = \int d^{4}x \,\, \{i b^{a}\partial^{\mu } A_{\mu }^{a} + \overline{c}^{a}\partial^{\mu }(D_{\mu }c)^{a}\}
\end{equation}
determines the action
\begin{equation}
 S = S_{YM+OP} + S_{gf}  
\end{equation}
which is left invariant under the following set of BRST transformations
\begin{eqnarray}\label{symmcmass}
sA_{\mu }^{a}&=& -(\partial _{\mu }c^{a}+gf^{abc}A_{\mu }^{b}c^{c})\nonumber \\
sc^{a}&=&\frac{g}{2}f^{abc}c^{b}c^{c}\nonumber \\
s\overline{\omega }_{\mu \nu }^{a} &=& \overline{\varphi}_{\mu \nu }^{a} + gf^{abc}c^{b}\overline{\omega }_{\mu \nu }^{c}\nonumber \\
s\overline{\varphi}_{\mu \nu }^{a} &=& gf^{abc}c^{b}\overline{\varphi}_{\mu \nu }^{c}\nonumber \\
s\varphi_{\mu \nu }^{a} &=& \omega _{\mu \nu }^{a}+gf^{abc}c^{b}\varphi_{\mu \nu }^{c} \nonumber \\
s\omega _{\mu \nu }^{a} &=& gf^{abc}c^{b}\omega _{\mu \nu }^{c}\nonumber \\
s\overline{\psi } &=& \overline{\phi } \nonumber \\
s\overline{\phi } &=& 0 \nonumber \\
s\phi &=& \psi \nonumber \\
s\psi &=& 0.
\end{eqnarray}

It is now necessary to say a few words about the scalar sector and spontaneous symmetry breaking. The action (\ref{lcma}) is invariant under a global $U(1)$ transformation which is 
\begin{eqnarray}
\overline{\phi } &\rightarrow & e^{i\Lambda }\overline{\phi }  \;\;\;\;\; \overline{\varphi }^{a}_{\mu\nu } \rightarrow e^{-i\Lambda }\overline{\varphi }^{a}_{\mu\nu } \nonumber \\
\phi &\rightarrow & e^{-i\Lambda }\phi  \;\;\;\;\; \varphi^{a}_{\mu\nu } \rightarrow e^{i\Lambda }\varphi^{a}_{\mu\nu } \nonumber \\
\overline{\psi } &\rightarrow & e^{i\Lambda }\overline{\psi }  \;\;\;\;\; \overline{\omega  }^{a}_{\mu\nu } \rightarrow e^{-i\Lambda }\overline{\omega  }^{a}_{\mu\nu } \nonumber \\
\psi &\rightarrow & e^{-i\Lambda }\psi  \;\;\;\;\; \omega ^{a}_{\mu\nu } \rightarrow e^{i\Lambda }\omega ^{a}_{\mu\nu }
\label{U1}
\end{eqnarray}
and these symmetrie, together with(\ref{symmcmass}), is enough to ensure that 
\begin{eqnarray}
V(\overline{\phi },\phi ;\overline{\psi },\psi ) &=&   V(\overline{\phi }\phi ;\overline{\psi }\psi ) \nonumber \\
V(\overline{\phi }\phi ;\overline{\psi }\psi ) &=& V(\overline{\phi }\phi - \overline{\psi }\psi ).
\end{eqnarray}
Looking to (\ref{lcma}) the potential 
\begin{equation}
  V(\overline{\phi }\phi - \overline{\psi }\psi )=- m^{2}(\overline{\phi }\phi  -\overline{\psi }\psi ) +\frac{\lambda }{2}(\overline{\phi }\phi  -\overline{\psi }\psi )^{2},
\end{equation}
admits non zero expectation value for the scalar field
\begin{eqnarray}
  \frac{\partial V}{\partial \phi }&=&0 \,\,\,\,\,\, \Rightarrow \,\,\,\,\,\,<\overline{\phi }\phi > = \frac{m^{2}}{\lambda }.
\end{eqnarray} 
Due to the $U(1)$ global symmetry it is necessary to take
\begin{equation}\label{defmu}
  <\overline{\phi }>=<\phi >=\frac{m}{\sqrt{\lambda }}=\mu.
\end{equation}
It is important to emphasize here that in order to obtain the localization $<\overline{\phi }>$  and $<\phi >$ always appear in the propagator and observables as a pair $<\overline{\phi }><\phi >$. 
The redefinitions $\overline{\phi }\rightarrow \overline{\phi }-<\overline{\phi }>$ and $\phi \rightarrow \phi -<\phi >$ are the only necessary requirement in order to obtain the localization and after the integration over the localizing fields the bilinear term is given by
\begin{equation}
 S_{mass} =\int d^{4}x \,\, \{  4 <\overline{\phi} ><\phi > A_{\mu }^{a}(\delta^{\mu \nu } -\frac{\partial^{\mu }\partial^{\nu }  }{\partial^{2} })A_{\nu }^{a}\} 
\end{equation}
which is the desired mass contribution for the free propagator.
In a future section we will discuss in detail the gauge propagator.
It is also important that is possible to use such mechanism to study certain topologically nontrivial solutions to the gauge field.
\subsection{Small comment about the symmetry breaking mechanism} 

It is clear that the action (\ref{lcma}) is constructed in the form:
\begin{equation}
S_{YM+OP} =\int d^{4}x \,\, \{  \frac{1}{4}F_{\mu \nu }^{a}F_{\mu \nu }^{a} + s(\Theta) \}, 
\end{equation}
where $\Theta$ is constructed with ultraviolet dimension $4$ and ghost number $-1$ with the localizing and gauge fields. In the symmetric phase $s(\Theta)$ is a trivial term and the cohomology property ensure that the action is pure Yang-Mills from the geometrical point of view. The situation in the broken phase is a little different. 
Let us look at the terms of interaction of the localizing fields and the curvature. They are obtained as 
\begin{eqnarray}
\Theta&=& \overline{\theta }_{\mu \nu \alpha \beta }(D^{\nu }\overline{\omega}^{\alpha \beta  })^{a}\theta^{\mu \sigma \lambda \rho }(D_{\sigma }\varphi_{\lambda \rho })^{a} \nonumber \\&-& i\overline{\psi}\varphi_{\mu \nu }^{a}\theta^{\mu \nu \alpha \beta}F_{\alpha \beta }^{a} + i \phi \overline{\omega }_{\mu \nu }^{a}\overline{\theta }^{\mu \nu \alpha \beta }F_{\alpha \beta }^{a}\nonumber \\
 s(\Theta) &=& \overline{\theta }_{\mu \nu \alpha \beta }(D^{\nu }\overline{\varphi}^{\alpha \beta  })^{a}\theta^{\mu \sigma \lambda \rho }(D_{\sigma }\varphi_{\lambda \rho })^{a}\nonumber \\
&-& \overline{\theta }_{\mu \nu \alpha \beta }(D^{\nu }\overline{\omega}^{\alpha \beta  })^{a}\theta^{\mu \sigma \lambda \rho }(D_{\sigma }\omega_{\lambda \rho })^{a}\nonumber \\
          &-& i\overline{\phi}\varphi_{\mu \nu }^{a}\theta^{\mu \nu \alpha \beta}F_{\alpha \beta }^{a} 
+ i\phi \overline{\varphi }_{\mu \nu }^{a}\overline{\theta }^{\mu \nu \alpha \beta }F_{\alpha \beta }^{a}\nonumber \\ 
&+& i\overline{\psi }\omega _{\mu \nu }^{a}\theta^{\mu \nu \alpha \beta}F_{\alpha \beta }^{a} 
+ i\psi  \overline{\omega }_{\mu \nu }^{a}\overline{\theta }^{\mu \nu \alpha \beta }F_{\alpha \beta }^{a}. \nonumber \\
\end{eqnarray}
In the broken phase the set of BRST symmetries for the scalar fields became:
\begin{eqnarray}
s \overline{\psi } &=& \overline{\phi }-\mu \nonumber \\
s \overline{\phi } &=& 0 \nonumber \\
s \phi &=& \psi \nonumber \\
s \psi &=& 0,
\end{eqnarray}
where $\mu$ is defined in (\ref{defmu}).
Applying now the same procedure to the localizing sector above, now we obtain one more term $i\mu\varphi_{\mu \nu }^{a}\theta^{\mu \nu \alpha \beta}F_{\alpha \beta }^{a}$. It is clear that doing the shift $\phi\Rightarrow \phi-\mu$ and $\overline{\phi}\Rightarrow\overline{\phi}-\mu$ in to the action another term is obtained. The term is $-i\mu \overline{\varphi }_{\mu \nu }^{a}\overline{\theta }^{\mu \nu \alpha \beta }F_{\alpha \beta }^{a}$ which is gauge-invariant. The relation of these term to the other terms of these sector is provided by a symmetry
\begin{equation}
\int d^{4}x \,\, \{ \overline{\varphi }_{\mu \nu }^{a}\frac{\delta S}{\delta\overline{\varphi }_{\mu \nu }^{a}} -\varphi_{\mu \nu }^{a}\frac{\delta S}{\delta\varphi_{\mu \nu }^{a}}+\overline{\phi}\frac{\delta S}{\delta\overline{\phi}}-\phi\frac{\delta S}{\delta\phi}\} =0,
\end{equation}
which can be easily extended to the broken phase and means that the U(1) symmetry presented in (\ref{U1}) is broken.
\begin{eqnarray}
&&\int d^{4}x \,\, \{ \overline{\varphi }_{\mu \nu }^{a}\frac{\delta S_{\mu}}{\delta\overline{\varphi }_{\mu \nu }^{a}} -\varphi_{\mu \nu }^{a}\frac{\delta S_{\mu}}{\delta\varphi_{\mu \nu }^{a}}+(\overline{\phi}-\mu)\frac{\delta S_{\mu}}{\delta\overline{\phi}}-(\phi-\mu)\frac{\delta S_{\mu}}{\delta\phi}\} =0 \nonumber \\
&&S_{\mu}= S(<\varphi>=\mu;<\overline{\varphi}>=\mu).
\end{eqnarray}
This equation does the link between the two phases and fixes the value of the mass in a similar way as used in the symmetry breaking mechanism. The main difference from these to the usual one is that mass  is for all the $A_{\nu}^{a}$ also the same that appears in Delbourgo-Thompson \cite{Delbourgo:1986wz,Delbourgo:1987np,Ruegg:2003ps} mechanism is avoided here.

\section{Extending to the quantum level}

The set of equations, valid at quantum level as any equation that obeys the quantum action principle (QAP) \cite{Piguet:1995er}, that ensure that no mixing term of the form $\overline{\varphi}_{\mu \nu }^{a}\varphi_{\mu \nu }^{a}$ is permited are
\begin{eqnarray}
 \theta_{\mu \nu \alpha \beta} \frac{\delta S}{\delta \overline{\varphi}_{\alpha \beta}^{a}} &=&0 \nonumber \\
 \overline{\theta }_{\mu \nu \alpha \beta} \frac{\delta S}{\delta \varphi_{\alpha \beta}^{a}} &=&0 \nonumber \\
 \theta_{\mu \nu \alpha \beta} \frac{\delta S}{\delta \overline{\omega }_{\alpha \beta}^{a}} &=&0 \nonumber \\
 \overline{\theta }_{\mu \nu \alpha \beta} \frac{\delta S}{\delta \omega_{\alpha \beta}^{a}} &=&0.
\end{eqnarray}
In order to present all equations compatible with the quantum action principle,  it is necessary to add to the action all the symmetries coupled to classical sources. It is interesting to remember here that due to the properties of the projectors it is useful to introduce the projectors explicitly into the source terms in a way that the projections over the source equations are also obtained. It is clear that introducing directly the projectors with the sources the functional derivative in respect to the sources does not give us only the symmetry  associated to these source but instead we obtain the projected symmetry which is also a symmetry of the action due to the operator idempotency . The Landau gauge fixing action plus the symmetries is given by:
\begin{eqnarray}
 S_{J} &=&\int d^{4}x \,\, \{ -\Omega^{a}(D_{\mu }C)^{a} \nonumber \\
 &+& L^{a}\frac{g}{2}f^{abc}c^{b}c^{c} + \overline{J}_{\alpha \beta }^{a}\theta ^{\alpha \beta \mu \nu }(\omega _{\mu \nu }^{a}+gf^{abc}c^{b}\varphi_{\mu \nu }^{c})\nonumber \\
 &+& J_{\alpha \beta }^{a}\overline{\theta }^{\alpha \beta \mu \nu }(gf^{abc}c^{b}\overline{\varphi}_{\mu \nu }^{c})
 + \chi_{\alpha \beta }^{a}\overline{\theta }^{\alpha \beta \mu \nu }(\overline{\varphi}_{\mu \nu }^{a} + gf^{abc}c^{b}\overline{\omega }_{\mu \nu }^{c}) \nonumber \\
 &+& \overline{\chi }_{\alpha \beta }^{a}\theta ^{\alpha \beta \mu \nu }(gf^{abc}c^{b}\omega _{\mu \nu }^{c}) \}.
\end{eqnarray}
Which has over the sources the same type of property as presented over the equations of motion for the fields
\begin{eqnarray}
 \overline{\theta}_{\mu \nu \alpha \beta} \frac{\delta S^{J} }{\delta \overline{J}_{\alpha \beta}^{a}} &=&0 \nonumber \\
 \theta_{\mu \nu \alpha \beta} \frac{\delta S^{J} }{\delta J_{\alpha \beta}^{a}} &=&0 \nonumber \\
 \overline{\theta}_{\mu \nu \alpha \beta} \frac{\delta S^{J} }{\delta \overline{\chi }_{\alpha \beta}^{a}} &=&0 \nonumber \\
 \theta_{\mu \nu \alpha \beta} \frac{\delta S^{J} }{\delta \chi _{\alpha \beta}^{a}} &=&0
\end{eqnarray}
Now the quantum actions and the Slavnov-Taylor identity are given by:
\begin{eqnarray}
\Sigma &=& S + S_{gf} + S_{J} \nonumber \\
S(\Sigma )&=& \int d^{4}x \,\,\{\frac{\delta \Sigma }{\delta A_{\mu }^{a}}\frac{\delta \Sigma }{\delta \Omega_{\mu }^{a}} + \frac{\delta \Sigma }{\delta c^{a}}\frac{\delta \Sigma }{\delta L^{a}} + \frac{\delta \Sigma }{\delta \overline{\varphi}_{\alpha \beta}^{a}}\frac{\delta \Sigma }{\delta J_{\alpha \beta}^{a}} \nonumber \\
&+& \frac{\delta \Sigma }{\delta \varphi_{\alpha \beta}^{a}}\frac{\delta \Sigma }{\delta \overline{J}_{\alpha \beta}^{a}}
+ \frac{\delta \Sigma }{\delta \overline{\omega }_{\alpha \beta}^{a}}\frac{\delta \Sigma }{\delta \chi_{\alpha \beta}^{a}} 
+ \frac{\delta \Sigma }{\delta \omega _{\alpha \beta}^{a}}\frac{\delta \Sigma }{\delta \overline{\chi }_{\alpha \beta}^{a}} \nonumber \\
&+& \overline{\phi } \frac{\delta \Sigma }{\delta \overline{\psi } } + \psi \frac{\delta \Sigma }{\delta \phi  } 
+ i b^{a}\frac{\delta \Sigma }{\delta \overline{c}^{a} }\} \nonumber \\
\end{eqnarray}
and the self-dual and anti-self-dual equations extended to the quantum action are:
\begin{eqnarray}
\theta_{\mu \nu \alpha \beta} \frac{\delta \Sigma }{\delta \overline{\varphi}_{\alpha \beta}^{a}} &=&0 \nonumber \\
 \overline{\theta }_{\mu \nu \alpha \beta} \frac{\delta \Sigma }{\delta \varphi_{\alpha \beta}^{a}} &=&0 \nonumber \\
 \theta_{\mu \nu \alpha \beta} \frac{\delta \Sigma }{\delta \overline{\omega }_{\alpha \beta}^{a}} &=&0 \nonumber \\
 \overline{\theta }_{\mu \nu \alpha \beta} \frac{\delta \Sigma }{\delta \omega_{\alpha \beta}^{a}} &=&0 
\nonumber \\
 \overline{\theta}_{\mu \nu \alpha \beta} \frac{\delta \Sigma }{\delta \overline{J}_{\alpha \beta}^{a}} &=&0 \nonumber \\
 \theta_{\mu \nu \alpha \beta} \frac{\delta \Sigma }{\delta J_{\alpha \beta}^{a}} &=&0 \nonumber \\
 \overline{\theta}_{\mu \nu \alpha \beta} \frac{\delta \Sigma }{\delta \overline{\chi }_{\alpha \beta}^{a}} &=&0 \nonumber \\
 \theta_{\mu \nu \alpha \beta} \frac{\delta \Sigma }{\delta \chi _{\alpha \beta}^{a}} &=&0.
\end{eqnarray}
It is important to enphasize again that this set of equations simply block the possibility of mass terms like $\overline{\varphi}_{\alpha \beta}^{a}\varphi^{\alpha \beta a}$ or $\overline{\omega }_{\alpha \beta}^{a} \omega^{\alpha \beta a}$ and thus turn the mass term obtained from the localization of the nonlocal operator presented in action and the symmetry breaking mechanism (\ref{lcma})into a stable one. Also this set of equations blocks a quartic term for the localing fields due to the simple fact that every antisymmetric tensor of rank $D+1$ in $D$ dimensions is zero then it is always possible to rewrite the indexes of a quadratic term in order to be forbidden by this set of equations.

The linearized operator $\beta_{\Sigma }$ is easily obtained from the Slavnov-Taylor identity and is given by:
\begin{eqnarray}
\beta_{\Sigma } &=& \int d^{4}x \,\, \{\frac{\delta \Sigma }{\delta A_{\mu }^{a}}\frac{\delta  }{\delta \Omega_{\mu }^{a}} + \frac{\delta \Sigma }{\delta \Omega_{\mu }^{a}}\frac{\delta  }{\delta A_{\mu }^{ai}}\nonumber \\
&+& \frac{\delta \Sigma }{\delta c^{a}}\frac{\delta  }{\delta L^{a}} + \frac{\delta \Sigma }{\delta L^{a}} \frac{\delta  }{\delta c^{a}} + \frac{\delta \Sigma }{\delta \overline{\varphi}_{\alpha \beta}^{a}}\frac{\delta  }{\delta J_{\alpha \beta}^{a}} + \frac{\delta \Sigma }{\delta J_{\alpha \beta}^{a}}\frac{\delta  }{\delta \overline{\varphi}_{\alpha \beta}^{a}}\nonumber \\
&+& \frac{\delta \Sigma }{\delta \varphi_{\alpha \beta}^{a}}\frac{\delta  }{\delta \overline{J}_{\alpha \beta}^{a}}
+ \frac{\delta \Sigma }{\delta \overline{J}_{\alpha \beta}^{a}}\frac{\delta }{\delta \varphi_{\alpha \beta}^{a}}
+ \frac{\delta \Sigma }{\delta \overline{\omega }_{\alpha \beta}^{a}}\frac{\delta  }{\delta \chi_{\alpha \beta}^{a}}
+ \frac{\delta \Sigma }{\delta \chi_{\alpha \beta}^{a}}\frac{\delta  }{\delta \overline{\omega }_{\alpha \beta}^{a}}\nonumber \\
&+& \frac{\delta \Sigma }{\delta \omega _{\alpha \beta}^{a}}\frac{\delta  }{\delta \overline{\chi }_{\alpha \beta}^{a}}
+ \frac{\delta \Sigma }{\delta \overline{\chi }_{\alpha \beta}^{a}}\frac{\delta  }{\delta \omega _{\alpha \beta}^{a}}
+ \overline{\phi } \frac{\delta  }{\delta \overline{\psi } } + \psi \frac{\delta  }{\delta \phi  } 
+ i b^{a}\frac{\delta  }{\delta \overline{c}^{a} }\}.
\end{eqnarray}
The gauge fixing, the antighost equation and the ghost equation that are characteristic from the Landau gauge are given by:\footnote{These set of equations ensure that the ghost fields do not renormalize in the Landau gauge} 
\begin{eqnarray}
  \frac{\delta \Sigma }{\delta b^{a}}&=&  i \partial^{\mu } A_{\mu }^{a}.\nonumber \\
  \frac{\delta \Sigma }{\delta \overline{c}^{a} }+ \partial _{\mu }\frac{\delta \Sigma }{\delta \Omega_{\mu }^{a} }&=&0. \nonumber \\
  G^{a}(\Sigma ) &=& \int d^{4}x \,\,\{ \frac{\delta \Sigma }{\delta c^{a}} + i g f^{abc}\overline{c}^{b}\frac{\delta \Sigma }{\delta b^{c}} \} \nonumber \\
  \Delta^{a} &=& \int d^{4}x \,\,g f^{abc}\{\Omega_{\mu }^{bi}A^{c \mu } - L^{b}c^{c} \nonumber \\
  &+& \theta_{\mu \nu \alpha \beta}(\overline{J}^{\mu \nu b} \varphi^{\alpha \beta c} - \overline{\chi} ^{\mu \nu b}\omega^{\alpha \beta c})\nonumber \\
  &+& \overline{\theta}_{\mu \nu \alpha \beta}(J^{\mu \nu b} \overline{\varphi}^{\alpha \beta c} - \chi^{\mu \nu b}\overline{\omega} ^{\alpha \beta c})  \} \nonumber \\
  G^{a}(\Sigma ) &=& \Delta^{a}.
\end{eqnarray}
The rigid equation that corresponds, in the Landau gauge, to the anticommutation of the ghost equation and the Slavnov one
\begin{eqnarray}
\{ G^{a},\beta_{\Sigma }\} &=& - W^{a}\nonumber \\
 W^{ai} &=& \int d^{4}x \,\,g f^{abc}\{A_{\mu }^{b}\frac{\delta }{\delta A_{\mu }^{c}} + \Omega _{\mu }^{b}\frac{\delta }{\delta \Omega _{\mu }^{c}} +  L^{b}\frac{\delta }{\delta L^{c}} + c^{b}\frac{\delta }{\delta c^{c}} \nonumber \\
 &+& b^{b}\frac{\delta }{\delta b^{c}} + \overline{c}^{b}\frac{\delta }{\delta \overline{c}^{c}}  \nonumber \\
 &+& \theta_{\mu \nu \alpha \beta}(\overline{J}^{\mu \nu b}\frac{\delta }{\delta \overline{J}_{\alpha \beta }^{c}}+ \overline{\chi }^{\mu \nu bi}\frac{\delta }{\delta \overline{\chi }_{\alpha \beta }^{c}}
 + \varphi^{\mu \nu b}\frac{\delta }{\delta \varphi _{\alpha \beta }^{c}} + \omega ^{\mu \nu b}\frac{\delta }{\delta \omega _{\alpha \beta }^{ci}} ) \nonumber \\
 &+& \overline{\theta}_{\mu \nu \alpha \beta}(J^{\mu \nu bi}\frac{\delta }{\delta \overline{J}_{\alpha \beta }^{c}}+ \chi^{\mu \nu b}\frac{\delta }{\delta \chi_{\alpha \beta }^{c}}
 + \overline{\varphi}^{\mu \nu b}\frac{\delta }{\delta \overline{\varphi }_{\alpha \beta }^{c}} + \overline{\omega }^{\mu \nu b}\frac{\delta }{\delta \overline{\omega }_{\alpha \beta }^{c}} )\}.  \nonumber \\
\end{eqnarray}
Also another set of equations that are compatible with the QAP are:
\begin{eqnarray}
 Q(\Sigma )&=&\int d^{4}x \{  \overline{\varphi}_{\mu \nu}^{ a}\frac{\delta \Sigma }{\delta \overline{\varphi }_{\mu \nu  }^{a}}-
 \varphi_{\mu \nu}^{ a}\frac{\delta \Sigma }{\delta \varphi_{\mu \nu  }^{a}} + \overline{J}_{\mu \nu}^{ a}\frac{\delta \Sigma }{\delta \overline{J }_{\mu \nu  }^{a}}-
 J_{\mu \nu}^{ a}\frac{\delta \Sigma }{\delta J_{\mu \nu  }^{a}} \nonumber \\
 &+&  \overline{\phi } \frac{\delta \Sigma }{\delta \overline{\phi } }-\phi \frac{\delta \Sigma }{\delta \phi }\} \nonumber \\
 \Delta &=&\int d^{4}x \{\overline{\theta}^{\mu \nu \alpha \beta}\chi_{\mu \nu }^{ai}\overline{\varphi }_{\alpha \beta }^{a} 
 + \theta^{\mu \nu \alpha \beta}\overline{J }_{\mu \nu  }^{a}\chi_{\alpha \beta }^{a}\} \nonumber \\
 Q(\Sigma )&=& \Delta
\label{Q}
\end{eqnarray}
\begin{eqnarray}
 R(\Sigma )&=&\int d^{4}x \{  \overline{\omega }_{\mu \nu}^{ a}\frac{\delta \Sigma }{\delta \overline{\omega  }_{\mu \nu  }^{a}}-
 \omega _{\mu \nu}^{ a}\frac{\delta \Sigma }{\delta \omega _{\mu \nu  }^{a}} + \overline{\chi }_{\mu \nu}^{ a}\frac{\delta \Sigma }{\delta \overline{\chi }_{\mu \nu  }^{a}}-
 \chi _{\mu \nu}^{ a}\frac{\delta \Sigma }{\delta \chi _{\mu \nu  }^{a}} \nonumber \\
 &+&  \overline{\psi  } \frac{\delta \Sigma }{\delta \overline{\psi } }-\psi \frac{\delta \Sigma }{\delta \psi }\} \nonumber \\ 
 R(\Sigma )&=& - \Delta 
\label{R}
\end{eqnarray}
The sum of equations (\ref{Q},\ref{R}) corresponds to a local implementation of the $U(1)$ (\ref{U1}) symmetry and corresponds to the quantum implementation of (\ref{U1}).

\begin{eqnarray}
[Q,\beta_{\Sigma } ]&=& \eta \nonumber \\
 \eta(\Sigma) &=& \int d^{4}x \{\overline{\phi } \frac{\delta \Sigma }{\delta \overline{\psi } }+\psi \frac{\delta \Sigma }{\delta \phi } + \overline{\theta}^{\mu \nu \alpha \beta}(\overline{\varphi }_{\mu \nu }^{a}\frac{\delta \Sigma }{\delta \overline{\omega }_{\alpha \beta }^{a} } - \chi_{\mu \nu }^{a}\frac{\delta \Sigma }{\delta J_{\alpha \beta }^{a}} )\nonumber \\
 &+& \theta^{\mu \nu \alpha \beta}(\omega _{\mu \nu }^{a}\frac{\delta \Sigma }{\delta \varphi _{\alpha \beta }^{a} } - \overline{J} _{\mu \nu }^{a}\frac{\delta \Sigma }{\delta \overline{\chi } _{\alpha \beta }^{a}} ) \} 
\end{eqnarray}
This set of equations, with the hermiticity condition, is enough to guarantee that no mass term for the localizing fields exists. Now it is necessary to guarantee that at the bilinear level the localizing action generates a mass term for the gauge fields $A_{\mu }^{a}$. In order to do that it is enough to take the bilinear action and integrate over the localizing fields.

\subsection{Bilinear sector of the localizing fields and the propagators}

Taking into account the bilinear part of the gauge functional with localizing fields and the gauge field it is possible to do the integration in the localizing fields
\begin{eqnarray}
Z[A,c,b]&=&\int DA_{\mu}\int D\varphi  D\overline{\varphi} e^{-S_{o}(\overline{\varphi  },\varphi ,A_{\mu})}\nonumber \\
S_{0(\overline{\varphi  },\varphi )} &=&\int d^{4}x \,\, \{\overline{\theta }_{\mu \nu \alpha \beta }(\partial ^{\nu }\overline{\varphi}^{\alpha \beta  })^{a}\theta^{\mu \sigma \lambda \rho }(\partial _{\sigma }\varphi_{\lambda \rho })^{a} \nonumber \\
 &-& ia\mu\varphi_{\mu \nu }^{a}\theta^{\mu \nu \alpha \beta}F_{0 \alpha \beta }^{a} 
 + ia\mu\overline{\varphi }_{\mu \nu }^{a}\overline{\theta }^{\mu \nu \alpha \beta }F_{0 \alpha \beta }^{a}\}\nonumber \\
 F_{0 \alpha \beta }^{a}&=& \partial_{\alpha }A_{\beta }^{a} - \partial_{\beta  }A_{\alpha }^{a},
\end{eqnarray}
where $<\overline{\phi}>=\mu$ and $<\phi >=\mu$ are the the vaccum obtained from a spontaneous symmetry breaking mechanism over the fields $\phi$ and $\overline{\phi}$
In order to do that integration it is enough to obtain the classical equations of motion which are:
\begin{eqnarray}
 -\{\overline{\theta }_{\mu \nu \alpha \beta }\theta^{\mu \sigma \lambda \rho }(\partial ^{\nu }\partial _{\sigma }\varphi_{\lambda \rho })^{a} \}+  ia\mu\overline{\theta }_{\mu \nu \alpha \beta }F^{0 a \mu\nu }&=& 0, \nonumber \\
 -\{\theta_{\mu \nu \alpha \beta }\overline{\theta}^{\mu \sigma \lambda \rho }(\partial ^{\nu }\partial _{\sigma }\overline{\varphi}_{\lambda \rho })^{a} \}-  ia\mu\theta_{\mu \nu \alpha \beta }F^{0 a\mu\nu }&=& 0. 
\end{eqnarray}
The classical solution for this set of equations of motion is of the form:
\begin{eqnarray}
  \varphi_{\mu \nu }^{a }&=& 4i a\mu \frac{1}{\partial^{2} }\theta_{\mu \nu \alpha \beta }F_{0}^{\alpha \beta a} \nonumber \\
  \overline{\varphi}_{\mu \nu }^{a }&=& -4i a\mu \frac{1}{\partial^{2} }\overline{\theta}_{\mu \nu \alpha \beta }F_{0}^{\alpha \beta a}.
\end{eqnarray}
Integrating over the localizing fields it is easy to find the contribution for the mass of the gauge field as:
\begin{equation}
 S_{mass} =\int d^{4}x \,\, \{  4(a)^{2} \mu^{2} A_{\mu }^{a}(\delta^{\mu \nu } -\frac{\partial^{\mu }\partial^{\nu }  }{\partial^{2} })A_{\nu }^{a}\}. 
\end{equation}
It is clear that $S_{mass}$ corresponds to the bilinear contribution to the gauge-invariant mass term presented in (\ref{lcma}) for the value $a=1$. It is important to emphasize here that the mass term is proportional to 
\begin{equation}
 \int d^{4}x \,\,\{A_{\mu}^{aT}A^{\mu aT}\},
\end{equation} where 
\begin{equation}
 A_{\mu}^{aT}=(\delta^{\mu \nu } -\frac{\partial^{\mu }\partial^{\nu }  }{\partial^{2} })A_{\nu}^{a}
\end{equation} is the transverse part of the gauge field. The complete nonabelian extension could be understood as the localization of $A^{2}_{min}$. These operator, defined as:
\begin{eqnarray}
 A^{2}_{min}&\equiv & min_{u}Tr\int d^{4}x \,\,\{A_{\mu}^{u}A^{\mu u}\}\nonumber \\
A_{\mu}^{u}&=&u^{\dagger}A_{\mu}u + \frac{i}{g}u^{\dagger}\partial_{\mu}u
\end{eqnarray}
and there relation to a non-Abelian nonlocal operator already has been obtained in Ref \cite{Capri:2005dy}.

Taking into account that we are working in the Landau gauge it is easy to obtain \footnote{We are using the convention that $<\Theta(-k)\Theta(k)>= - G_{\Theta\Theta}$}:
\begin{equation}
 <A^{a}_{\mu }(-k) A^{b}_{\nu }(k)>=-2\delta^{ab}  (\delta^{\mu \nu } -\frac{k^{\mu }k^{\nu }  }{k^{2} })\frac{1}{k^{2}+8(a)^{2}\mu^{2}},
\end{equation}
performing some calculations we obtain for all the propagators
\begin{eqnarray}
<\overline{\varphi}_{\mu \nu }^{a } (-k) A^{b}_{\gamma }(k)>&=&-\delta^{ab}\frac{2}{a\mu}(\frac{1}{k^{2}}- \frac{1}{k^{2}+8(a)^{2}\mu^{2}})
\overline{\theta}_{\mu\nu\gamma\sigma}k^{\sigma}\nonumber \\
<\varphi_{\mu \nu }^{a } (-k) A^{b}_{\gamma }(k)>&=&\delta^{ab}\frac{2}{a\mu}(\frac{1}{k^{2}}- \frac{1}{k^{2}+8(a)^{2}\mu^{2}})
\theta_{\mu\nu\gamma\sigma}k^{\sigma}\nonumber \\
<\varphi_{\sigma\lambda}^{a}(k) \overline{\varphi}_{\mu \nu }^{b}(-k)>&=&\delta^{ab}\frac{2}{k^{2}+8(a)^{2}\mu^{2}}
(\theta_{\sigma\lambda\rho\epsilon}k^{\epsilon}\overline{\theta}_{\mu\nu\alpha\gamma}k^{\gamma}\delta^{\rho\alpha})(\frac{1}{k^{^{2}}})\nonumber \\
<\omega_{\sigma\lambda}^{a}(k)\overline{\omega}_{\mu \nu }^{b}(-k)>&=&-\delta^{ab}\frac{2}{k^{2}}
(\theta_{\sigma\lambda\rho\epsilon}k^{\epsilon}\overline{\theta}_{\mu\nu\alpha\gamma}k^{\gamma}\delta^{\rho\alpha})(\frac{1}{k^{^{2}}})\nonumber \\
<\overline{\phi}(-k)\phi(k)>&=&-\frac{1}{2}(\frac{1}{k^{2}+2\lambda\mu^{2}} +\frac{1}{k^{2}})\nonumber \\
<\overline{\phi}(-k)\overline{\phi}(k)>&=&<\phi(-k)\phi(k)>=-\frac{1}{2}(\frac{1}{k^{2}+2\lambda\mu^{2}} -\frac{1}{k^{2}})\nonumber \\
<\overline{\psi}(-k)\psi(k)>&=&\frac{1}{k^{2}}\nonumber \\
<b^{a}(-k)A^{b}_{\nu }(k)>&=&-\delta^{ab}\frac{k_{\nu}}{k^{2}}\nonumber \\
<\overline{c}^{a}(-k)c^{b}(k)>&=&\frac{\delta^{ab}}{k^{2}}.
\end{eqnarray}
This set of propagators confirms our assumption that this mechanism generates a transverse mass term for the gauge field. It is also possible to observe the non massive poles in $<\overline{\varphi}_{\mu \nu }^{a } (-k) A^{b}_{\gamma }(k)>$ and $<\varphi_{\mu \nu }^{a } (-k) A^{b}_{\gamma }(k)>$ expected from the symmetry breaking mechanism. Moreover. It is relevant to emphasize again that by construction the model is renormalizable due to the geometrical preoperties of the tensorial fields. Finally one can note that the mixing terms between the tensorial terms and the gauge curvature take into account the values of a scalar field and can be usefull to study topological properties of these action but this is a very extended task and certainly demands another work.

\section{Conclusions}
In this work we present a possible extension of the main idea presented in Ref \cite{Capri:2005dy} that is to localize a non-Abelian gauge-invariant operator in order to obtain gauge-invariant mass term. The method uses a symmetry breaking mechanism and obtain the same mass to all components of the gauge field. This can be useful in order to study the operator $A^{2}_{mim}$ that is quite important in many aspects of confinement \cite{Semenov,Zwanziger:1990tn,Dell'Antonio:1989jn,Dell'Antonio:1991xt}. We present all the necessary equations that are compatible to the quantum action principle that can be used to prove the renormalizability of the model. We have obtained the important property that mass terms in the localizing fields are blocked by the geometrical properties of these fields while maintaining the localizing property. We
point out that this action possesses a small number of parameters, a feature that is useful for higher order computations.

The possibility of having at our disposal a true local and renormalizable action might provide us with a consistent framework for a future investigation of the possible implications of nonlocal gauge-invariant operators of ultraviolet dimension two.

\section*{ Acknowledgments} 

This study was financed in part by the Coordena\c{c}\~{a}o de Aperfei\c{c}oamento de Pessoal de N\'{\i}vel Superior - Brasil (Capes) - Finance Code 001. The Conselho Nacional de Desenvolvimento Cient\'{\i}fico e Tecnol\'{o}gico (CNPq-Brazil) and the SR2-UERJ are gratefully acknowledged for financial support. M. A. L. Capri is a level PQ-2 researcher under the program Produtividade em Pesquisa-CNPq, 302040/2017-0.


\begin{thebibliography}{999}\bibitem{Cornwall:1981zr}  J.~M.~Cornwall,
%``Dynamical Mass Generation In Continuum QCD,''
Phys.\ Rev.\ D \textbf{26}, 1453 (1982).

%\cite{Greensite:1985vq}

\bibitem{Greensite:1985vq}  J.~Greensite and M.~B.~Halpern,
%``Variational Computation Of Glueball Masses In Continuum QCD,''
Nucl.\ Phys.\ B \textbf{271}, 379 (1986).

%\cite{Stingl:1985hx}

\bibitem{Stingl:1985hx}  M.~Stingl,
%``Propagation Properties And Condensate Formation Of The Confined Yang-Mills
%Field,''
Phys.\ Rev.\ D \textbf{34}, 3863 (1986) [Erratum-ibid.\ D
\textbf{36}, 651 (1987)].

%\cite{Lavelle:1988eg}

\bibitem{Lavelle:1988eg}  M.~J.~Lavelle and M.~Schaden,
%``Propagators And Condensates In QCD,''
Phys.\ Lett.\ B \textbf{208}, 297 (1988).

%\cite{Gubarev:2000nz}

\bibitem{Gubarev:2000nz}  F.~V.~Gubarev and V.~I.~Zakharov,
%``On the emerging phenomenology of <(A(a)(mu))**2(min)>,''
Phys.\ Lett.\ B \textbf{501}, 28 (2001) [arXiv:hep-ph/0010096].

%\cite{Gubarev:2000eu}

\bibitem{Gubarev:2000eu}  F.~V.~Gubarev, L.~Stodolsky and V.~I.~Zakharov,
%``On the significance of the quantity A**2,''
Phys.\ Rev.\ Lett.\ \textbf{86}, 2220 (2001)
[arXiv:hep-ph/0010057].
%%CITATION = HEP-PH 0010057;%%

\bibitem{Verschelde:2001ia}  H.~Verschelde, K.~Knecht, K.~Van Acoleyen and
M.~Vanderkelen,
%``The non-perturbative groundstate of QCD and the local composite  operator
%A(mu)**2,''
Phys.\ Lett.\ B \textbf{516}, 307 (2001) [arXiv:hep-th/0105018].

%\cite{Kondo:2001nq}

\bibitem{Kondo:2001nq}  K.~I.~Kondo,
%``Vacuum condensate of mass dimension 2 as the origin of mass gap and  quark
%confinement,''
Phys.\ Lett.\ B \textbf{514}, 335 (2001) [arXiv:hep-th/0105299].
%%CITATION = HEP-TH 0105299;%%

%\cite{Kondo:2001tm}

\bibitem{Kondo:2001tm}  K.~I.~Kondo, T.~Murakami, T.~Shinohara and T.~Imai,
%``Renormalizing a BRST-invariant composite operator of mass dimension 2  in
%Yang-Mills theory,''
Phys.\ Rev.\ D \textbf{65}, 085034 (2002) [arXiv:hep-th/0111256].
%%CITATION = HEP-TH 0111256;%%

%\cite{Dudal:2003vv}

\bibitem{Dudal:2003vv}  D.~Dudal, H.~Verschelde, R.~E.~Browne and
J.~A.~Gracey,
%``A determination of  and the non-perturbative vacuum energy of
%Yang-Mills theory in the Landau gauge,''
Phys.\ Lett.\ B \textbf{562}, 87 (2003) [arXiv:hep-th/0302128].
%%CITATION = HEP-TH 0302128;%%

%\cite{Browne:2003uv}

\bibitem{Browne:2003uv}  R.~E.~Browne and J.~A.~Gracey,
%``Two loop effective potential for  in the Landau gauge in  quantum
%chromodynamics,''
JHEP \textbf{0311}, 029 (2003) [arXiv:hep-th/0306200].
%%CITATION = HEP-TH 0306200;%%

%\cite{Dudal:2003gu}

\bibitem{Dudal:2003gu}  D.~Dudal, H.~Verschelde, V.~E.~R.~Lemes,
M.~S.~Sarandy, S.~P.~Sorella and M.~Picariello,
%``Gluon-ghost condensate of mass dimension 2 in the Curci-Ferrari gauge,''
Annals Phys.\ \textbf{308}, 62 (2003) [arXiv:hep-th/0302168].
%%CITATION = HEP-TH 0302168;%%

%\cite{Dudal:2003by}

\bibitem{Dudal:2003by}  D.~Dudal, H.~Verschelde, J.~A.~Gracey,
V.~E.~R.~Lemes, M.~S.~Sarandy, R.~F.~Sobreiro and S.~P.~Sorella,
%``Dynamical gluon mass generation from  in linear covariant gauges,''
JHEP \textbf{0401}, 044 (2004) [arXiv:hep-th/0311194].
%%CITATION = HEP-TH 0311194;%%

%\cite{Dudal:2004rx}

\bibitem{Dudal:2004rx}  D.~Dudal, J.~A.~Gracey, V.~E.~R.~Lemes,
M.~S.~Sarandy, R.~F.~Sobreiro, S.~P.~Sorella and H.~Verschelde,
%``An analytic study of the off-diagonal mass generation for Yang-Mills theories
%in the maximal Abelian gauge,''
Phys.\ Rev.\ D \textbf{70}, 114038 (2004) [arXiv:hep-th/0406132].

%\cite{Browne:2004mk}

\bibitem{Browne:2004mk}  R.~E.~Browne and J.~A.~Gracey,
%``One loop MS-bar gluon pole mass from the LCO formalism,''
Phys.\ Lett.\ B \textbf{597}, 368 (2004) [arXiv:hep-ph/0407238].
%%CITATION = HEP-PH 0407238;%%

%\cite{Gracey:2004bk}

\bibitem{Gracey:2004bk}  J.~A.~Gracey,
%``Two loop MS-bar gluon pole mass from the LCO formalism,''
Eur.\ Phys.\ J.\ C \textbf{39}, 61 (2005) [arXiv:hep-ph/0411169].
%%CITATION = HEP-PH 0411169;%%

%\cite{Li:2004te}

\bibitem{Li:2004te}  X.~d.~Li and C.~M.~Shakin,
%``Description of gluon propagation in the presence of an A**2 condensate,''
Phys.\ Rev.\ D \textbf{71}, 074007 (2005) [arXiv:hep-ph/0410404].
%%CITATION = HEP-PH 0410404;%%

%\cite{Boucaud:2001st}

\bibitem{Boucaud:2001st}  P.~Boucaud, A.~Le Yaouanc, J.~P.~Leroy,
J.~Micheli, O.~Pene and J.~Rodriguez-Quintero,
%``Testing Landau gauge OPE on the lattice with a  condensate,''
Phys.\ Rev.\ D \textbf{63}, 114003 (2001) [arXiv:hep-ph/0101302].
%%CITATION = HEP-PH 0101302;%%

%\cite{Boucaud:2002nc}

\bibitem{Boucaud:2002nc}  P.~Boucaud, J.~P.~Leroy, A.~Le Yaouanc, J.~Micheli, O.~Pene, F.~De Soto, A.~Donini, H.~Moutare and J.~Rodriguez-Quintero
%``Instantons and  condensate,''
Phys.\ Rev.\ D \textbf{66}, 034504 (2002) [arXiv:hep-ph/0203119].
%%CITATION = HEP-PH 0203119;%%

%\cite{Boucaud:2005rm}

\bibitem{Boucaud:2005rm}  P.~Boucaud, F.~de Soto, J.~P.~Leroy, A.~Le Yaouanc, J.~Micheli, H.~Moutarde,
O.~Pene and J.~Rodriguez-Quintero, %``Artefacts and  power corrections: Revisiting the MOM Z(psi)(p**2) and
%Z(V),''
arXiv:hep-lat/0504017. %%CITATION = HEP-LAT 0504017;%%

%\cite{RuizArriola:2004en}

\bibitem{RuizArriola:2004en}  E.~Ruiz Arriola, P.~O.~Bowman and
W.~Broniowski,
%``Landau-gauge condensates from the quark propagator on the lattice,''
Phys.\ Rev.\ D \textbf{70}, 097505 (2004) [arXiv:hep-ph/0408309].
%%CITATION = HEP-PH 0408309;%%

%\cite{Suzuki:2004dw}

\bibitem{Suzuki:2004dw}  T.~Suzuki, K.~Ishiguro, Y.~Mori and T.~Sekido,
%``The dual Meissner effect and Abelian magnetic displacement currents,''
Phys.\ Rev.\ Lett.\ \textbf{94}, 132001 (2005)
[arXiv:hep-lat/0410001].
%%CITATION = HEP-LAT 0410001;%%

%\cite{Gubarev:2005it}

\bibitem{Gubarev:2005it}  F.~V.~Gubarev and S.~M.~Morozov,
%`` condensate, Bianchi identities and chromomagnetic fields  degeneracy
%in SU(2) YM theory,''
Phys.\ Rev.\ D \textbf{71}, 114514 (2005) [arXiv:hep-lat/0503023].
%%CITATION = HEP-LAT 0503023;%%

%\cite{Furui:2005bu}

\bibitem{Furui:2005bu}  S.~Furui and H.~Nakajima,
%``Infrared features of KS fermion and Wilson fermion in lattice Landau gauge
%QCD,''
arXiv:hep-lat/0503029. %%CITATION = HEP-LAT 0503029;%%

%\cite{Boucaud:2005xn}

\bibitem{Boucaud:2005xn}  P.~Boucaud, J.~P.~Leroy, A.~Le Yaouanc, A.~Y.~Lokhov, J.~Micheli, O.~Pene,
J.~Rodriguez-Quintero and C.~Roiesnel,
%``Non-perturbative power corrections to ghost and gluon propagators,''
arXiv:hep-lat/0507005. %%CITATION = HEP-LAT 0507005;%%

%\cite{Chernodub:2005gz}

\bibitem{Chernodub:2005gz}  M.~N.~Chernodub, K.~Ishiguro, Y.~Mori, Y.~Nakamura, M.~I.~Polikarpov,
T.~Sekido, T.~Suzuki and V.~I.~Zakharov,
%``Vacuum type of SU(2) gluodynamics in maximally Abelian and Landau gauges,''
arXiv:hep-lat/0508004. %%CITATION = HEP-LAT 0508004;%%

%\cite{Dudal:2003gu}

\bibitem{Dudal:2003gu}  D.~Dudal, H.~Verschelde, V.~E.~R.~Lemes,
M.~S.~Sarandy, S.~P.~Sorella and M.~Picariello,
%``Gluon-ghost condensate of mass dimension 2 in the Curci-Ferrari gauge,''
Annals Phys.\ \textbf{308}, 62 (2003) [arXiv:hep-th/0302168].
%%CITATION = HEP-TH 0302168;%%
%\cite{Dudal:2003np}

\bibitem{Dudal:2003np}  D.~Dudal, H.~Verschelde, V.~E.~R.~Lemes,
M.~S.~Sarandy, R.~F.~Sobreiro, S.~P.~Sorella and J.~A.~Gracey,
%``Renormalizability of the local composite operator A(mu)**2 in linear
%covariant gauges,''
Phys.\ Lett.\ B \textbf{574}, 325 (2003) [arXiv:hep-th/0308181].
%%CITATION = HEP-TH 0308181;%%

%\cite{Dudal:2003pe}

\bibitem{Dudal:2003pe}  D.~Dudal, H.~Verschelde, V.~E.~R.~Lemes, M.~S.~Sarandy, R.~F.~Sobreiro,
S.~P.~Sorella, M.~Picariello and J.~A.~Gracey,
%``The anomalous dimension of the gluon-ghost mass operator in Yang-Mills
%theory,''
Phys.\ Lett.\ B \textbf{569}, 57 (2003) [arXiv:hep-th/0306116].
%%CITATION = HEP-TH 0306116;%%

%\cite{Delbourgo:1986wz}
\bibitem{Delbourgo:1986wz} 
  R.~Delbourgo and G.~Thompson,
  %``Massive, Unitary, Renormalizable Yang-mills Theory Without Higgs Mesons,''
  Phys.\ Rev.\ Lett.\  {\bf 57}, 2610 (1986).

%\cite{Delbourgo:1987np}
\bibitem{Delbourgo:1987np} 
  R.~Delbourgo, S.~Twisk and G.~Thompson,
  %``Massive Yang-mills Theory: Renormalizability Versus Unitarity,''
  Int.\ J.\ Mod.\ Phys.\ A {\bf 3}, 435 (1988).

%\cite{Ruegg:2003ps}
\bibitem{Ruegg:2003ps} 
  H.~Ruegg and M.~Ruiz-Altaba,
  %``The Stueckelberg field,''
  Int.\ J.\ Mod.\ Phys.\ A {\bf 19}, 3265 (2004)

\bibitem{Capri:2005dy} 
  M.~A.~L.~Capri, D.~Dudal, J.~A.~Gracey, V.~E.~R.~Lemes, R.~F.~Sobreiro, S.~P.~Sorella and H.~Verschelde,
  %``A Study of the gauge invariant, nonlocal mass operator Tr integer d**4 x F mu nu (D**2)**-1 F mu nu in Yang-Mills theories,''
  Phys.\ Rev.\ D {\bf 72}, 105016 (2005) [hep-th/0510240].

%\cite{Piguet:1995er}
\bibitem{Piguet:1995er} 
  O.~Piguet and S.~P.~Sorella,
  %``Algebraic renormalization: Perturbative renormalization, symmetries and anomalies,''
  Lect.\ Notes Phys.\ Monogr.\  {\bf 28}, 1 (1995).

\bibitem{Semenov}  Semenov-Tyan-Shanskii and V.A. Franke, Zapiski Nauchnykh
Seminarov Leningradskogo Otdeleniya Matematicheskogo Instituta im.
V.A. Steklov AN SSSR, Vol. \textbf{120} (1982) 159. English
translation: New York: Plenum Press 1986.

%\cite{Zwanziger:1990tn}

\bibitem{Zwanziger:1990tn}  D.~Zwanziger,
%``Quantization Of Gauge Fields, Classical Gauge Invariance And Gluon
%Confinement,''
Nucl.\ Phys.\ B \textbf{345}, 461 (1990). %%CITATION = NUPHA,B345,461;%%

%\cite{Dell'Antonio:1989jn}

\bibitem{Dell'Antonio:1989jn}  G.~Dell'Antonio and D.~Zwanziger,
%``Ellipsoidal Bound On The Gribov Horizon Contradicts The Perturbative
%Renormalization Group,''
Nucl.\ Phys.\ B \textbf{326}, 333 (1989). %%CITATION = NUPHA,B326,333;%%

%\cite{Dell'Antonio:1991xt}

\bibitem{Dell'Antonio:1991xt}  G.~Dell'Antonio and D.~Zwanziger,
%``Every gauge orbit passes inside the Gribov horizon,''
Commun.\ Math.\ Phys.\ \textbf{138}, 291 (1991).
%%CITATION = CMPHA,138,291;%%

%\cite{vanBaal:1991zw}

\bibitem{vanBaal:1991zw}  P.~van Baal,
%``More (thoughts on) Gribov copies,''
Nucl.\ Phys.\ B \textbf{369}, 259 (1992). %%CITATION = NUPHA,B369,259;%%




\end{thebibliography}
\end{document}